\documentclass{INTERSPEECH2023}
\usepackage{amsmath,graphicx}
\usepackage{xcolor, amssymb}
\usepackage{caption}
\usepackage{subcaption}
\usepackage{enumitem}
\interspeechcameraready

\setlist{nosep}

\title{Learning When to Trust Which Teacher for Weakly Supervised ASR}
\name{Aakriti Agrawal$^*$\thanks{Work done while the first author was an intern with Amazon Alexa.}
\quad Milind Rao$^\dagger$ \quad Anit Kumar Sahu$^\dagger$ \quad Gopinath Chennupati$^\dagger$ \quad Andreas Stolcke$^\dagger$}
\address{$^*$University of Maryland, U.S.A. \quad $^\dagger$Amazon Alexa AI, U.S.A.}
\email{agrawal5@umd.edu,\{milinrao,anitsah,chennug,stolcke\}@amazon.com}
\begin{document}
\ninept
\maketitle
\begin{abstract}
Automatic speech recognition (ASR) training can utilize multiple experts as teacher models, each trained on a specific domain or accent. Teacher models may be opaque in nature since their architecture may be not be known or their training cadence is different from that of the student ASR model. Still, the student models are updated incrementally using the pseudo-labels generated independently by the expert teachers. In this paper, we exploit supervision from multiple domain experts in training student ASR models. This training strategy is especially useful in scenarios where few or no human transcriptions are available. To that end, we propose a {\em Smart-Weighter} mechanism that selects an appropriate expert based on the input audio, and then trains the student model in an unsupervised setting. We show the efficacy of our approach using LibriSpeech and LibriLight benchmarks and find an improvement of 4 to 25\% over baselines that uniformly weight all the experts,
 use a single expert model, or combine experts using ROVER.

\end{abstract}
%
\noindent\textbf{Index terms:} ASR, teacher-student training, semi-supervised learning, self-supervised learning, ROVER.
%
\section{Introduction}
\label{sec:intro}
Self-supervised learning approaches~\cite{thomas2013deep,kahn2020self,chennupati2022ilasr} for ASR usually rely on a single expert teacher model to generate pseudo-labels to train the student ASR models. ROVER \cite{fiscus1997post} is a classic technique to generate a single best transcription by aligning alternate teacher hypotheses and using a rule like majority voting. Mixture-of-expert (MoE) approaches for speech~\cite{salinasknowledge,student-teacher}, on the other hand, make use of the MoE layers in training the student model. Generally, when we have multiple independent experts, each of them is trained to be performant for a specific domain. In practice, these expert models can not be typically deployed on devices with limited compute and storage. Even in a resource-rich, cloud-based setting, these experts are hard to train because the experts are heterogeneous in terms of their structure (e.g., hybrid or deep neural networks based), in terms of their dependence on external language models, and in terms of size. Therefore, we treat these experts as opaque generators of transcripts for a given audio input.




As an alternative to using MoE layers, which involves additional access to the experts beyond output transcripts, we propose {\em Smart-Weighter}, a method that selects a domain expert among many and with access limited to the generated transcripts. The selection of experts is conditioned on the input audio, i.e., for a given training utterance we select the best expert for generating the teacher transcript.


In this paper, we use the streaming-compatible recurrent neural network transducer (RNN-T)~\cite{RNNT} ASR model, whose training objective is to maximize the probability of the transcript tokens given the audio and the past context. We develop three RNN-T-based domain experts and a separate student model, all of which are of the same size. We train the experts on LibriSpeech; the experts are trained on mutually exclusive data subsets to mimic  domain experts. The student model is then trained, along with the {\em Smart-Weighter} network, on untranscribed audio, with the selected experts producing the transcripts. While this framework is generally applicable to alternate forms of feedback, such as weak supervision, in this work we focus on using transcripts from expert models. 

Our main contribution is an unsupervised framework for learning from multiple expert models using a {\em Smart-Weighter} network that selects domain experts based on the unlabeled input audio.

\vspace{-4mm}
\section{Related Work}

ROVER \cite{fiscus1997post} combines multiple transcripts using equal weights or recognition confidences \cite{audhkhasi2013empirical} only. Smart-Weighter additionally makes use of utterance audio to determine transcript relevance. Another disadvantage of ROVER is that it asymptotes quickly as  the number of experts increases. Furthermore, a ROVER expert can itself be added to the list of experts that the Smart-Weighter weights over. Alternatively, Smart-Weighter could be used to estimate the weights of the different inputs prior to combination with ROVER.

A popular approach to teacher-student learning is knowledge distillation (KD). In KD, the student is trained to match with the teacher's output distribution by minimizing the KL divergence between either their bottleneck layer or output activations. However, KD is not practical with opaque experts that do not provide access to activations. Although KD was originally used to learn from a single model, \cite{multiKD} uses multiple teachers while using uncertainty-based KD. 
In \cite{RLmultiKD} reinforcement learning (RL) is used to select teachers for KD. The approach relies on feedback/reward based on the performance of the student model to update the policy parameters.
Unlike in \cite{RLmultiKD}, here we consider an unsupervised setting where we do not have access to ground-truth transcripts to generate rewards, such as word error rate~(WER). KD also has been used for multi-level-multi-teacher \cite{Adap} and methods based on error rate instead of loss \cite{DistK}. 

Another line of work utilizing multiple experts is mixture of experts (SpeechMoe2 \cite{speechmoe2}, DeepMoe \cite{deepmoe}). A router chooses among experts in each neural network layer during training, given an input acoustic embedding. In this setting, the experts and student are trained end-to-end, leading to a less computationally expensive solution when expert and student are combined. \cite{RLmoe} uses an RL-based policy to mask the activation in each layer in the mixture of experts. This method is not applicable to our scenario since our domain experts are typically large and reside in the cloud. Also, the experts cannot be trained along with the student model. 

Speech enhancement solutions have also been proposed to use multiple experts. Like our work, \cite{adaptive} presents a gating mechanism for expert selection and loss propagation in end-to-end training. The gating mechanism learns to choose appropriate weights while training the experts. \cite{adaptive2} uses an ensemble model to train specialized experts based on different speakers, giving significant gains in speech enhancement. Speakers are partitioned based on their characteristics, using k-means clustering. We took inspiration from the clustering-based approach and gating mechanism found in this earlier work.

\vspace{-2mm}
\section{Method}
\vspace{-2mm}

\begin{figure}[tb]
  \centering
  \includegraphics[width=0.75\linewidth,trim={0cm 0cm 0cm 0cm},clip]{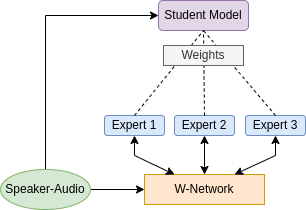}
\caption{Visualization of training or updating a student model given unlabeled audio. For a given utterance, we have teacher transcripts from multiple opaque experts of differing quality. A Smart-Weighter (W-network) consumes expert transcriptions and utterance audio to weight their quality, with a larger weight given to experts deemed to be more accurate. The student model is trained using semi-supervised learning with audio and paired expert transcriptions using the determined weights.}
\label{fig:method_visualization}
\end{figure}

Our focus is on training and updating student ASR models given unlabeled audio and arbitrary opaque teacher ASR models. In this section, we first describe the Smart-Weighter model that either selects an appropriate expert or weights its transcripts for a given utterance. We then describe the setup used to obtain teacher models using LibriSpeech \cite{LS}, although the ideas developed here are applicable to arbitrary experts. Finally, we describe how student ASR models are trained from scratch given a stream of unlabeled audio (e.g., LibriLight \cite{LL}) using transcripts obtained from multiple experts weighted by the Smart-Weighter model. This overall workflow is shown in Figure \ref{fig:method_visualization}.  

\subsection{Smart-Weighter}
\label{sec:smart-weighter}

In order to train a student model using these multiple expert transcriptions of differing quality paired with audio, we develop a Smart-Weighter
that selects or weights an appropriate expert given the utterance input. 

This is done by generating weights for each expert that sum to $1$, thereby conforming the weighting to a probability simplex. A larger weight is assigned to experts that are deemed to be more accurate for that utterance.

\begin{figure}[bt]
  \centering
  \includegraphics[width=1\linewidth,trim={0cm 0cm 0cm 0cm},clip]{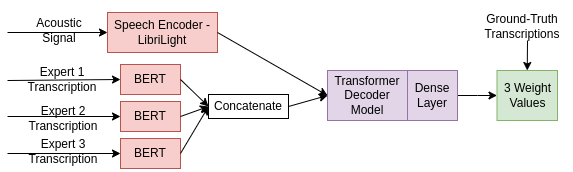}
\caption{The Smart-Weighter consists of a speech encoder that produces features from an utterance audio and a BERT language model that produces features from expert transcriptions. A transformer-decoder model consumes the BERT features while cross-attending to audio features. The outputs are processed to determine the weights of the expert models.}
\label{fig:W_network}
\vspace{-0.5cm}
\end{figure}

The Smart-Weighter network shown in Figure \ref{fig:W_network} takes as input an acoustic signal and transcriptions from the expert models (we use three experts here, but the method is applicable to an arbitrary number). It uses a unidirectional LSTM-based speech encoder trained on the LibriLight dataset to generate acoustic signal embeddings and a pretrained BERT model \cite{devlin2018bert} to generate expert-transcription embeddings. The transcription embeddings are then joined together using a separator token embedding and used as input for a transformer-decoder model \cite{transformer} that cross-attends to the acoustic embeddings. We use a $6$-layer transformer-decoder model (512 units and 8 attention heads) that uses full self-attention on the input, which is a concatenation of the BERT embeddings of expert transcriptions with full cross-attention to relevant sections of the acoustic embeddings. The output of the transformer-decoder layer is then pooled and passed through feed-forward dense layers with intermediate ReLU activations and a final layer using softmax activation. Finally, we obtain weight values $w_i, i=1,2,3$, corresponding to the three experts. 

The Smart-Weighter is trained on a $100$-hour subset of the LibriSpeech dataset also using ground-truth transcriptions. For an utterance, we obtain expert weights $w_i$ and we develop target labels $z\in \{0, 1\}^3$ where $z_i=1$ if expert $i$ has the lowest word error rate (WER) as measured using ground-truth transcriptions, and $z_i=0$ otherwise. We then apply binary cross entropy loss, i.e., $L=-\sum_i z_i\log w_i +  (1- z_i)\log (1 - w_i)$ on each of the expert weights. Thus the Smart-Weighter is trained to upweight expert transcriptions that show lowest WER and produce lower weights for experts that show poor performance. 



\vspace{-2mm}
\subsection{Expert Setup}
\vspace{-2mm}

We treat the expert models as opaque models  that may have arbitrary architectures, model sizes, training methodologies or training sets, and may include unspecified domain-specific auxiliary language models. The experts may have comparable performance or be trained on complementary data with minimal overlap. It is unreasonable to expect an expert to be the best performer across all domains and for all utterances. Different expert models may outperform others depending on domain or context, especially when they have similar capacities or sizes. 

In order to simulate the variability of real-world expert ASR models, we trained ASR expert models using alternate splits of the $960$-hour LibriSpeech dataset. We train three experts using two different speaker partitioning strategies. The first method randomly partitions the speakers of the training set; the second method clusters the training speakers using k-means on their audio features (speaker embeddings). The sizes of the speaker partitions created by these two schemes (Random and Clustered) are shown in Table~\ref{tab:speaker_split}. We  expect the experts to perform similarly when trained on random speaker partitions,  and to be more complementary when trained on clusters based on speaker similarity. 

\begin{table}[t]
\centering
\caption{Number of speakers used for training expert ASR models on LibriSpeech partitions. Speakers are partitioned either randomly or clustered by speaker embeddings.} \label{tab:speaker_split}
 \resizebox{0.65\columnwidth}{!}{%
\begin{tabular}{|c|c|c|c|}
  \hline
  & Expert 1 & Expert 2 & Expert 3\\
 \hline
Random   & 779 & 779 & 780  \\
Clustered   & 488 & 1074 & 758 \\
 \hline
\end{tabular}
}
\end{table}

\vspace{-2mm}
\subsubsection{ASR Model}
\vspace{-2mm}

We use the recurrent neural network-transducer (RNN-T) architecture for all expert teacher models. We believe our method generalizes to different architectures and model sizes, but leave a study beyond RNN-Ts to future work. One advantage of this choice is that all experts can be deployed on resource-constrained devices which may not be feasible if larger architectures are used. 

The models have $60$M parameters with a $5\times1024$ LSTM encoder, a $2\times1024$ LSTM prediction network and a feed-forward joint network with $\textrm{tanh}$ activation. The input embeddings of the prediction network are $512$-dimensional. We use a $2500$ word-piece tokenizer. SpecAugment is used for the audio features. The audio features comprise $64$-dimensional log-Mel filter-bank energy features that are computed over a $25$-ms window with a $10$-ms shift. The features computed on three consecutive $10$-ms frames are stacked and subsampled to result in a $192$-dimensional features at a $30$-ms frame rate, which form the input to the ASR model. All expert and student models are trained to convergence (30-50K steps on 24 V100 GPUs) with Adam optimizer and learning rate of $10^{-5}$. 

\vspace{-2mm}
\subsubsection{Speaker Clustering}
\vspace{-2mm}

To partition speakers by similarity, we use a trained speaker identification model trained on LibriSpeech.\footnote{Specifically, we use the ResCNN model trained with triplet loss, as available on https://github.com/philipperemy/deep-speaker.}
For each speaker and utterance, we prepare mean- and covariance-normalized features for segments of length $1$ second. The speaker embedding for the utterance is the average of $10$ such evaluations of the embedding model. For each speaker, we obtain $10$ embeddings (for $10$ different randomly chosen utterances) that we make use of in the clustering procedure described below. 

\begin{figure}[tb]
\begin{subfigure}{.5\columnwidth}
  \includegraphics[scale=0.22]{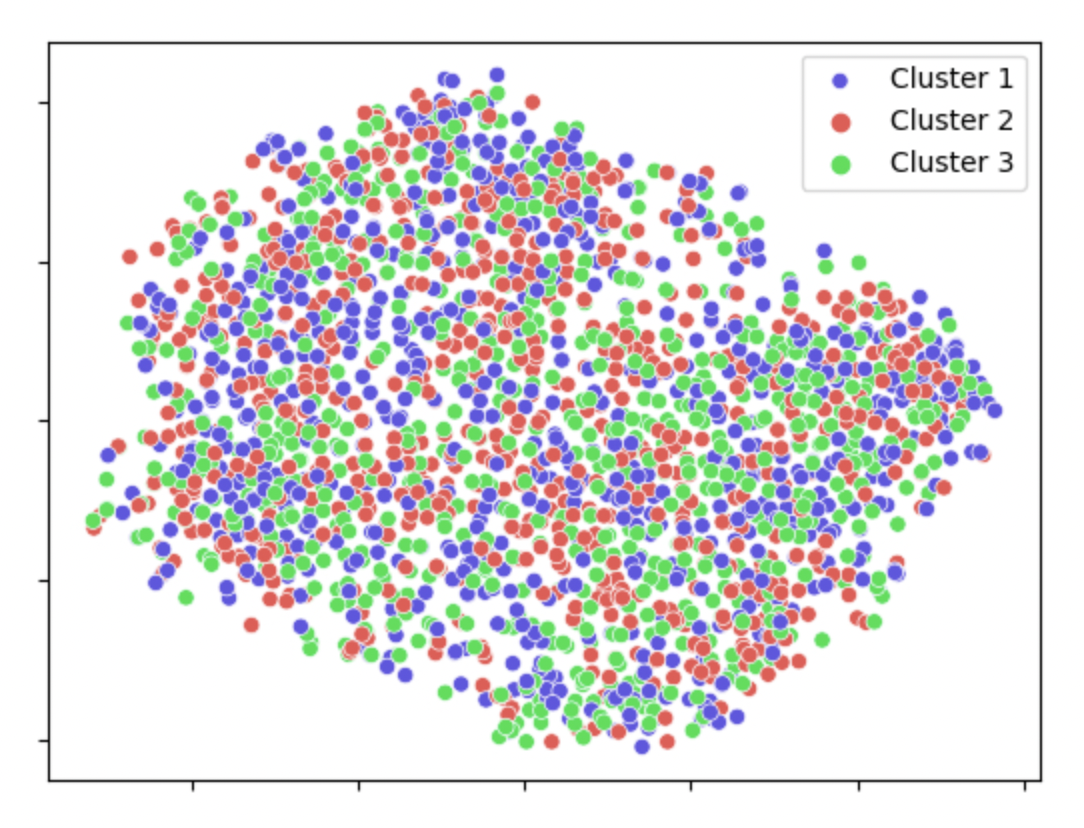}
\end{subfigure}
\begin{subfigure}{.1\columnwidth}
  \includegraphics[scale=0.22]{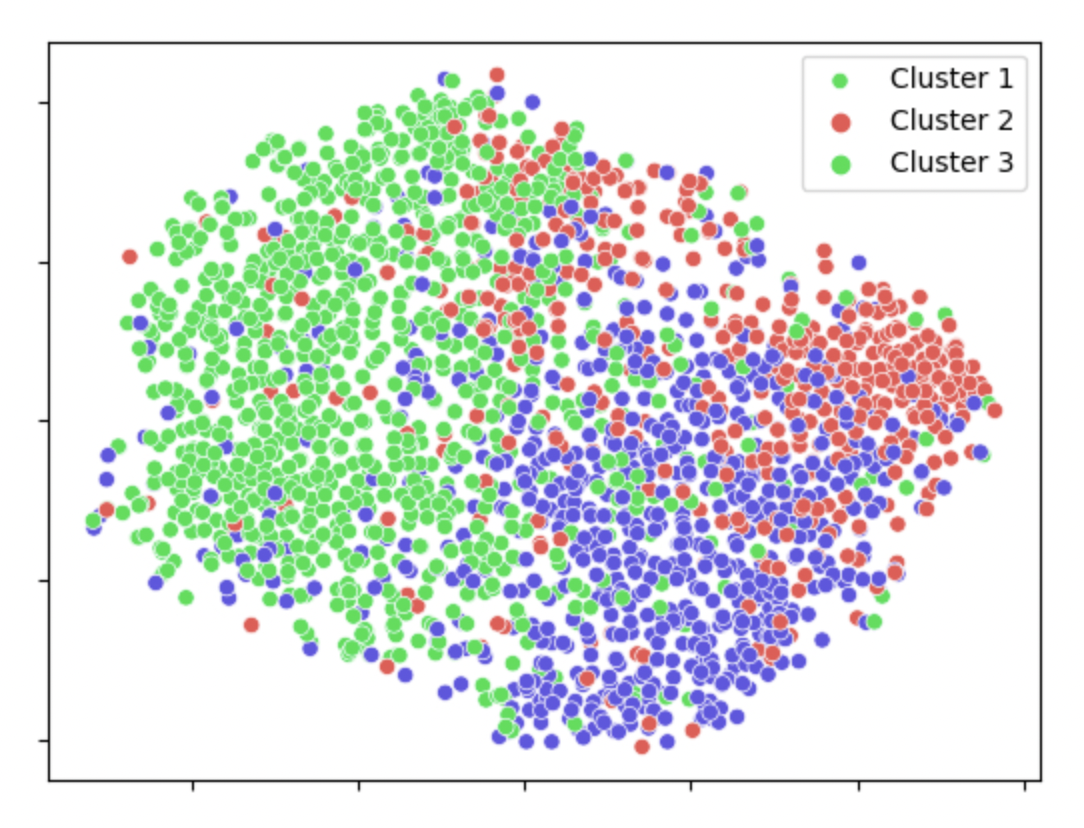}
\end{subfigure}
\caption{Speaker cluster assignments for expert ASR training based on random assignment (left) and speaker embedding based clusters (right).}
\label{fig:Clustering}
\end{figure}

We applied k-means clustering with a fixed initial state on the speaker embeddings to group them into three disjoint clusters.
We use a majority vote to assign a speaker to one of the three clusters based on their 10 embeddings.%
\footnote{As anecdotal validation, when using just two clusters, we obtained a partition that strongly correlated with gender annotations.} Figure \ref{fig:Clustering} shows the speaker cluster assignments for the random and clustered partitioning, using a t-SNE \cite{van2008visualizing} visualization to map the $512$-dimensional speaker embeddings to two dimensions.

\vspace{-2mm}
\subsection{Student Model}
\label{sec:student_model}
\vspace{-1mm}

Our student ASR model is an RNN-T similar to the experts, trained from scratch on unlabeled audio using the expert outputs as teacher. We employ a similar architecture for student and experts based on past findings \cite{student-teacher} that contrast model complexity and type of students and teachers and shows that most effective training occurs when model architectures are similar. Additionally, student and teacher models can run on resource-constrained devices. However, our methods are generally applicable to settings with differing model complexities for student and teacher models.  

Given a pair of audio features $x$ and transcription $t$, an RNN-T model is trained by minimizing the RNN-T loss $L_{\mathrm{RNNT}}(x,t)$ that maximizes the probability of obtaining transcription $t$ given $x$. The method used to combine the expert transcriptions $t_i$ impacts student performance. The Smart-Weighter described in Section~\ref{sec:smart-weighter} weights the expert transcripts based on inferred accuracy. We contrast this method with two baselines: 
\begin{itemize}[leftmargin=*]
    \item \textbf{Baseline 1: Best-Expert} is to train the student model using the transcription from a single best expert $t^*$ as determined from the validation set, i.e., the student model is trained with the loss function $L(x) = L_{\mathrm{RNNT}}(x, t^*)$.
    \item \textbf{Baseline 2: All-Experts} is to train the student model by weighting the transcriptions from all experts equally, i.e., the loss function for each utterance is the sum of loss functions for each of the experts: $L(x) = \frac{1}{3} \sum_i L_{\mathrm{RNNT}}(x, t_i)$.
\end{itemize}
We compare these baselines and ROVER combined expert transcription with our proposed method:
\begin{itemize}[leftmargin=*]
    \item \textbf{Smart-Weighter} produces weights $w_i$ corresponding to each of the expert transcriptions for an utterance. The weights $w_i$ produced have low entropy, i.e., the weight for one of the experts is close to 1 and others near 0. In order to make use of all available information, we flatten this distribution by renormalizing the weights using softmax with a temperature parameter ($T=1$), giving normalized weights $\hat{w}_i = \frac{e^{w_i/T}}{\sum_j e^{w_j/T}}$. Finally, the loss function for training the student model using the expert transcriptions weighted by the Smart-Weighter is $L(x) = \sum_i \hat{w}_i L_{\mathrm{RNNT}}(x, t_i).$
\end{itemize}


\section{Results}

\vspace{-2mm}
\subsection{Evaluating the Experts}
\vspace{-2mm}

Table \ref{tab:expert} shows the ASR performance of the experts on various LibriSpeech data splits. We can observe overfitting on the specific data split an expert is trained on (e.g., Expert $1$ on Expert $1$'s training split), but not on the other splits. We observe some variation in the performance of the experts obtained by speaker clustering, since, unlike for the random splits, these experts had differing numbers of utterances and speakers for training.
\begin{table}[t]
\caption{WER evaluation of experts on LibriSpeech data splits. The results for the choice of ``Best-Expert'' are highlighted.}
\centering
  \resizebox{\columnwidth}{!}{%
\begin{tabular}{|l||c|c|c||c|c|c|}
 \hline
 &  \multicolumn{3}{|c||}{Random Expert} &  \multicolumn{3}{|c|}{Clustered Expert}\\
 \hline
  & 1 & 2 & 3 & 1 & 2 & 3 \\
 \hline
TestClean   & 10.91 & 11.81 &  10.82 & 13.66 & 11.27 & 15.94 \\
TestOther   & 25.56 & 26.53 &  25.27 & 30.5 & 28.57 & 35.18 \\
DevClean &	10.77 & 11.45 & \bf 10.78 &	13.65 & \bf 10.99 & 16.07 \\
DevOther & 25.18 & 25.29 & \bf 24.27 &	29.75 &	2\bf 7.02 & 33.16 \\
Train &	9.74 &	9.93 &	9.24 &	12.08 &	8.48 &	16.15 \\
Train split & & & & & & \\
\quad Expert-1 & 0.34 & 14.28 & 13.29 & 0.14 & 15.65 & 20.1 
\\
\quad Expert-2 & 14.32 & 0.74 &14.14 & 18.09 & 0.13 & 20.68
\\
\quad Expert-3 & 14.25 & 14.97 & 0.38 & 17.18 & 15.36 & 0.18
\\
 \hline
\end{tabular}
}
\label{tab:expert}
\end{table}

\vspace{-2mm}
\subsection{Student Model Evaluation}
\vspace{-2mm}

The student model is trained on $10$K-hour subset of the LibriLight dataset and evaluated on LibriSpeech train, dev and test splits.
For the Best-Expert baseline we chose Expert~3 among the random experts, and Expert~2 for the clustered experts, based on the highlighted results shown in Table~\ref{tab:expert} and train a student model.
The All-Experts baseline model is trained by giving all experts equal weight, as described in Section~\ref{sec:student_model}. As another baseline, we also evaluate using ROVER \cite{fiscus1997post} to combine teacher transcripts. 

Table~\ref{tab:student} shows all results when training the student on LibriLight data. ROVER outperforms the best teacher model comparing with Table~\ref{tab:entropy}. Distilling a student on a large dataset makes it better than the corresponding teacher model (comparing Best-Expert student model to the best performing teacher). All-Experts performs better than Best-Expert and ROVER, showing the value of multiple experts.
For random experts,
Smart-Weighter shows an improvement of $4\%$ and $3\%$ compared to All-Experts, $25\%$ and $18\%$ compared to Best-Expert, and $20\%$ and $15\%$ compared to ROVER baseline on test-clean and test-other splits. For clustered experts, we see $4\%$ and $5\%$ improvement compared to All-Experts, $13\%$ and $12\%$  compared to Best-Expert, $26\%$ and $23\%$ on test-clean and test-other.  Best-Expert is more competitive with clustered experts, possibly because the best expert also has the largest training set.

Table~\ref{tab:student2} shows results with a student model when trained on LibriSpeech data, but without using ground-truth transcriptions. Here we can also study what an ``oracle'' expert could achieve, i.e., choosing the expert with the most accurate output for each training utterance, giving us a bound on the performance of an expert-weighting model.   With such an oracle, we see larger improvements for clustered experts than for random experts; this could be simply because clustered experts have more variation in their output quality.

Notably, for different datasets as well as different configurations of expert models, using a smart weighter produces statistically significant improvement compared to the baselines.
We also note the counter-intuitive finding that the end-to-end student ASR evaluation with random experts is better than with clustered experts. We suppose this is because of the unequal cluster sizes in the clustered case. We defer the investigation of optimal cluster assignment for expert training to future work.

\begin{table}[t]

\caption{WER results with {\bf student model trained on LibriLight and experts trained on LibriSpeech}, using different teacher configurations and training speaker partitioning methods (Random and Clustered).}
\centering
  \resizebox{\columnwidth}{!}{%
\begin{tabular}{|c|c|c|c|c|c|}
 \hline
  Teacher & TestClean & TestOther  & Train & DevClean & DevOther\\
  \hline
& \multicolumn{5}{c|}{Random} \\
 \hline
 ROVER Baseline & 9.32 & 22.84 & \textbf{7.62} & 9.52 & 21.90 \\ \hline
Best-Expert  & 10.03 & 23.69 & 11.80   & 9.89 & 22.83 \\
All-Experts   &  7.79 & 19.93 & 8.56 & 7.72 &  19.46 \\
\textbf{Smart-Weighter} & \textbf{7.47} & \textbf{19.37} & 7.95 &  \textbf{7.34} & \textbf{18.85}  \\
 \hline
 & \multicolumn{5}{c|}{Clustered} \\
 \hline
 ROVER Baseline & 11.08 & 27.55 & \textbf{9.06} & 10.97 & 26.34 \\ \hline
Best-Expert   & 9.40 & 24.09 & 11.07 & 9.23 & 23.62\\
All-Experts   & 8.55 & 22.33 & 9.81 & 8.42 & 21.53\\
\textbf{Smart-Weighter} &  \textbf{8.21} & \textbf{21.22} & 9.32 & \textbf{7.90} & \textbf{20.86} \\
 \hline
\end{tabular}
}
\label{tab:student}
\end{table}

\begin{table}[t]

\caption{
WER results with {\bf student and experts trained on LibriSpeech}, using different teacher and oracle configurations and training speaker partitioning methods (Random and Clustered).}
\centering
  \resizebox{\columnwidth}{!}{%
\begin{tabular}{|c|c|c|c|c|c|}
 \hline
 Teacher  & TestClean & TestOther  & Train & DevClean & DevOther\\
 \hline
 & \multicolumn{5}{c|}{Random} \\
 \hline
  ROVER Baseline & 9.32 & 22.84 & 7.62 & 9.52 & 21.90 \\ \hline
Best-Expert  & 10.15 & 23.45 & 9.99   & 10.19 & 22.70 \\
All-Experts   & 7.94  & 18.86 & 6.83 & 7.67 & 18.42  \\
\textbf{Smart-Weighter} & \textbf{7.53} & \textbf{18.93} & \textbf{5.30} & \textbf{7.22}  &  \textbf{18.17} \\
Oracle & 7.41 & 18.24 & 1.80 &  7.06 & 17.68   \\
 \hline
  & \multicolumn{5}{c|}{Clustered} \\
 \hline
 ROVER Baseline & 11.08 & 27.55 & 9.06 & 10.97 & 26.34 \\ \hline
Best-Expert   & 9.08  & 23.05 & 7.78 & 8.89 & 21.97  \\
All-Experts   & 8.33  & 20.92 & 6.95 & 8.16 & 20.35  \\
\textbf{Smart-Weighter} & \textbf{7.95} & \textbf{19.85} & \textbf{5.09} &  \textbf{7.70} & \textbf{19.29}  \\
Oracle & 7.08 & 17.88 & 1.82 &  6.83 & 17.48  \\
 \hline

\end{tabular}
}
\label{tab:student2}
\end{table}

\vspace{-2mm}
\subsection{Smart-Weighter Evaluation}
\vspace{-2mm}

We evaluate our Smart-Weighter using two metrics. The first, \textbf{accuracy}, is based on the percentage of transcriptions selected from the best teacher, i.e.,\ 100\% implies that the Smart-Weighter always assigned highest weight to the expert with lowest WER. Another metric, \textbf{weighted WER}, uses Smart-Weighter output to compute a weighted average of WERs $\sum_{i=1}^{3} w_i \cdot \mathrm{WER}(t^*, t_i)$. The intuition behind this metric is that, as the model learns, the weights should increase for lower-WER transcriptions and weighted WER will decrease.  
As shown in Table~\ref{tab:w_network}, the Smart-Weighter performs marginally better at associating acoustic profiles with relevant experts using clustered experts than with random experts .  

\begin{table}[t]
\caption{Smart-Weighter evaluation results on LibriSpeech}
\centering
  \resizebox{\columnwidth}{!}{%
\begin{tabular}{|c||c|c||c|c|}
 \hline
 &  \multicolumn{2}{|c||}{Random} &  \multicolumn{2}{|c|}{Clustered}\\
 \hline
  & Accuracy & Weighted WER  & Accuracy & Weighted WER  \\
 \hline
TestClean   & 60\% & 0.118 & 63\% & 0.134  \\
TestOther   & 49\% & 0.280  & 59\% & 0.315 \\
DevClean    & 61\% & 0.110  & 64\% & 0.130 \\
DevOther    & 50\% & 0.254 & 58\% & 0.286 \\
 \hline
\end{tabular}
}
\label{tab:w_network}
\end{table}

\subsubsection{Smart-Weighter with ASR entropy}

As a possible variant for expert weighting, we study the effect of adding a simple form of ASR confidence as side information to the W-network. This makes additional assumptions about the expert models, i.e., that $n$-best hypotheses and their scores are available. We compute the posterior hypothesis probabilities $p_i$ from their score values $s_i$, by normalization: $p_i = \frac{s_i}{\sum_j s_j}$.  The entropy is then computed as $H = -\sum_i p_i \log p_i.$
We use the $n = 10$ best hypotheses.

This entropy measure is low when the $1$-best hypothesis is assigned a score vastly higher than the other hypotheses, and higher when the ASR model is less confident in its best hypothesis. This entropy measure is injected into the Smart-Weighter model in the feed-forward layers before the final weights are obtained. Results seen in Table~\ref{tab:entropy} show a 3 to 4\% improvement in accuracy as compared to Table~\ref{tab:w_network} for random experts.  


\begin{table}[ht]

\caption{Smart-Weighter evaluation results on LibriSpeech splits after including ASR entropy information}
\centering
  \resizebox{\columnwidth}{!}{%
\begin{tabular}{|c||c|c||c|c|}
 \hline
 &  \multicolumn{2}{|c||}{Random} &  \multicolumn{2}{|c|}{Clustered}\\
 \hline
  & Accuracy & Weighted WER  & Accuracy & Weighted WER  \\
 \hline
TestClean   & 63\% & 0.113 & 64\% & 0.133  \\
TestOther   & 53\% & 0.274 & 60\% & 0.314 \\
DevClean    & 64\% & 0.109 & 64\% & 0.130 \\
DevOther    & 54\% & 0.244 & 57\% & 0.287 \\
 \hline
\end{tabular}
}
\label{tab:entropy}
\end{table}

\vspace{-2mm}
\section{Conclusions}
We have shown how to train student ASR models given unlabeled audio using teacher output from multiple opaque expert ASR models. An application of this framework is in continual adaptation of deployed ASR systems, using unlabeled audio and domain-specific experts. We have developed a Smart-Weighter that consumes audio features and expert transcriptions and upweights  experts that are deemed to be more accurate for a given training utterance. We simulated opaque ASR experts, with or without complementarity, using multiple speaker partitioning strategies. The student model trained with weighted expert teacher transcriptions showed a 4 to 25\% improvement over baselines that  weight all experts uniformly, choose a single expert, or combine transcriptions with ROVER.   
We also observed an improvement in the Smart-Weighter by using ASR confidence or entropy as an additional feature.

\noindent\footnotesize{\textbf{Acknowledgments:} We thank Anirudh Raju, Gautam Tiwari, Guruprasad Ramesh, Bach Bui, and Sri Subramaniam among many others at Alexa AI for helpful discussions.}


\bibliographystyle{plain}
\bibliography{references}

\end{document}